\documentclass[a4paper,11pt]{article}
\pdfoutput=1 

\usepackage{jheppub} 

\usepackage[T1]{fontenc} 

\usepackage{graphicx}
\graphicspath{{figures/}{fig/}}
\usepackage{amsmath}
\usepackage{bm}
\usepackage{slashed}
\usepackage{epsfig}
\usepackage{amsfonts}
\usepackage{epstopdf}
\usepackage{color}
\usepackage{extarrows}
\allowdisplaybreaks

\def\a{\alpha}
\def\b{\beta}

\def\d{\delta}
\def\e{\epsilon}
\def\g{\gamma}
\def\s{\sigma}
\def\th{\theta}

\newcommand\vq{\vec{\textbf{q}}}
\newcommand\abq{|\vec{\textbf{q}}|}

\def\cbar{{\overline{c}}}
\newcommand{\jpsi}{{J/\psi}}

\newcommand{\state}[4]{{^#1\hspace{-0.6mm}#2_{#3}^{[#4]}}}

\newcommand\CScSa{\state{3}{S}{1}{1}}

\newcommand\COaSz{\state{1}{S}{0}{8}}

\newcommand\COcSa{\state{3}{S}{1}{8}}
\newcommand\COcPz{\state{3}{P}{0}{8}}

\newcommand\COcPj{\state{3}{P}{J}{8}}

\newcommand\mo{{\mathcal O}}

\newcommand{\LDME}[2]{\langle\mo^{#1}(#2)\rangle}
\newcommand\mops{\LDME{\jpsi}{\CScSa}}
\newcommand\mopa{\LDME{\jpsi}{\COaSz}}
\newcommand\mopb{\LDME{\jpsi}{\COcSa}}
\newcommand\mopc{\LDME{\jpsi}{\COcPz}}
\newcommand\mopj{\LDME{\jpsi}{\COcPj}}

\newcommand{\LDMEn}[1]{\langle\mo^{#1}_n\rangle}
\newcommand\mopn{\LDMEn{\jpsi}}

\newcommand{\pLDME}[2]{\langle{\mathcal P}^{#1}(#2)\rangle}
\newcommand\pops{\pLDME{\jpsi}{\CScSa}}
\newcommand\popa{\pLDME{\jpsi}{\COaSz}}
\newcommand\popb{\pLDME{\jpsi}{\COcSa}}

\newcommand\lrD{\overleftrightarrow{\boldsymbol{D}}}

\newcommand\COcP{\state{3}{P}{}{8}}
\newcommand\mopP{\LDME{\jpsi}{\COcP}}
\newcommand\popP{\pLDME{\jpsi}{\COcP}}

\newcommand\mopN{\LDME{\jpsi}{n}}
\newcommand\popN{\pLDME{\jpsi}{n}}
\newcommand\cc{c\bar{c}}

\title{Relativistic effect of $J/\psi$ hadroproduction in large $p_T$ region}

\author[a]{Rong Li}
\author[b,c]{An-Ping Chen}
\author[b]{Jing-Kai Huang}
\author[b,c,d]{Yan-Qing Ma}

\affiliation[a]{School of Science, Xi'an Jiaotong University, Xi'an 710049,	China}
\affiliation[b]{School of Physics and State Key Laboratory of Nuclear Physics and Technology, Peking University,\\Beijing 100871, China}
\affiliation[c]{Center for High Energy Physics, Peking University,\\Beijing 100871, China}
\affiliation[d]{Collaborative Innovation Center of Quantum Matter,\\Beijing 100871, China}

\emailAdd{rongliphy@xjtu.edu.cn}
\emailAdd{chenanping@pku.edu.cn}
\emailAdd{jkhuangphysics@hotmail.com}
\emailAdd{yqma@pku.edu.cn}

\abstract{
By combining NRQCD factorization and collinear factorization, we
compute a series of relativistic corrections for $J/\psi$ hadroproduction
to all orders in $v^2$ at large $p_T$ limit. The $v^2$ expansion converges well
for all channels. We find that the ratio
of relativistic correction term to the corresponding leading term is independent
of kinematic variables for any channel, which generalizes the proportional relations found in previous works to all orders.
}

\begin{document}
\maketitle

\section{Introduction}\label{sec:introduction}

The study  of heavy quarkonium production is important to understand
hadronization physics of QCD. Due to the successful performance of the Tevatron and
the LHC, the $\jpsi$ hadroproduction receives especial attention recently.
The most widely used theory to describe heavy quarkonium
production in the past two decades is the non-relativistic QCD (NRQCD)
factorization~\cite{Bodwin:1994jh}. However, based on it, prediction
of prompt $\jpsi$ hadroproduction at leading order (LO) in $\a_s$
is significantly transversely polarized~\cite{Braaten:1999qk} which
contradicts with the measurement by the CDF
Collaboration~\cite{Affolder:2000nn,Abulencia:2007us}, where one
found that the produced $\jpsi$ is almost unpolarized and even
slightly longitudinally polarized as $p_T$ increases.

To solve the polarization puzzle, the complete next-to-leading order
(NLO) in $\alpha_s$ corrections for all important channels ($\CScSa$, $\COaSz$,
$\COcSa$ and $\COcPj$) have been calculated~\cite{Campbell:2007ws, Artoisenet:2007xi,
Gong:2008sn, Gong:2008hk, Ma:2010vd, Ma:2010yw,
Butenschoen:2010rq,Li:2011yc, Butenschoen:2012px, Chao:2012iv,
Gong:2012ug, Wang:2012is, Gong:2013qka, Han:2014kxa, Butenschoen:2014dra, Han:2014jya, Zhang:2014ybe, Bodwin:2015iua, Feng:2018ukp}. Full NLO prediction can indeed describes both yield
and polarization of prompt $\jpsi$ hadroproduction
consistently~\cite{Chao:2012iv,Bodwin:2014gia}, but the corresponding color-octet
(CO) long-distance matrix elements (LDMEs) are much larger than that
needed in $\jpsi$ production at B factories~\cite{Ma:2008gq,
Gong:2009kp, Zhang:2009ym}, which challenges the universality of
LDMEs. 
It seems like that still some other important contributions have not been
included yet in $\jpsi$ hadroproduction. A possible source may be
relativistic effects, as the relative momentum between $\cc$ pair is
not very small in $\jpsi$ system ($v^2\approx0.3$). Along this line, a soft gluon factorization was proposed to resum relativistic corrections to all orders \cite{Ma:2017xno}. In this paper, we will concentrate on relativistic corrections within NRQCD factorization. For $J/\psi$ hadroproduction, NLO relativistic
corrections are calculated in Refs.~\cite{Fan:2009zq,
Xu:2012am}, where it was found that, although corrections for
$\CScSa$ channel are small~\cite{Fan:2009zq}, corrections for CO
channels are non-ignorable~\cite{Xu:2012am}. As NLO relativistic
corrections are important, it is also needed to study the importance
of even higher order relativistic corrections to explore all important
contributions for $\jpsi$ hadroproduction and to test the
convergence of relativistic expansion. 
Another interesting
finding in Refs.~\cite{Fan:2009zq, Xu:2012am} is that, in large
transverse momentum $p_T$ limit, the ratio of relativistic
correction term to the corresponding leading term is a constant number for each of the
four channels. The proportional relation for $\CScSa$ channel is well
understood~\cite{Ma:2012ex}, but not for other three channels. An
understanding of these proportional relations is helpful to illustrate
the structure of relativistic effects. 

In high $p_T$ region, it is more convenient to use collinear factorization
framework~\cite{Nayak:2005rt, Nayak:2005rw, Kang:2011mg, Fleming:2012wy, Fleming:2013qu,Kang:2014tta,Kang:2014pya},  which has been rigorously proven to leading power (LP)
and next-to-leading power (NLP) in $p_T$ expansion \cite{Kang:2014tta}. As contributions
beyond NLP are surely negligible when $p_T$ is sufficiently large,
this collinear factorization method can capture the main feature in
this region. For example, using the collinear factorization, a simple LO calculation can
already reproduce NLO NRQCD calculations at large $p_T$~\cite{Ma:2014svb}. 
As the proportional relations mentioned in the last paragraph are
large $p_T$ behavior, they should also be reproducible in the collinear
factorization. In fact, we will see that, in large $p_T$ limit,
relativistic corrections for all the three CO channels can be easily calculated
to all orders in $v^2$ in the framework of collinear factorization
and the proportional relations hold to all orders.

The rest of the paper is organized as follows. In
Sec.~\ref{sec:NRQCD}, we review the NRQCD factorization and give the
explicit large $p_T$ relations that were found in
Refs.~\cite{Fan:2009zq, Xu:2012am}. We then review the collinear
factorization for heavy quarkonium production in
Sec.~\ref{sec:coll}. The combination of NRQCD factorization and
collinear factorization is also discussed. Based on these
factorizations, we study the relativistic effects of CO $\jpsi$
production via $\COcSa$, $\COaSz$ and $\COcP$ channels in
Sec.~\ref{sec:CORC} one by one. We summarize our study in
Sec.~\ref{sec:summary}. Finally, we provide formulas
to project short distance coefficients to definite orbital
angular momentum state in Appendix~\ref{sec:app}.

\section{Relativistic corrections within NRQCD
factorization}\label{sec:NRQCD}

In NRQCD factorization, differential cross section of the $\jpsi$
production at hadron colliders is factorized as~\cite{Bodwin:1994jh}
\begin{align}\label{eq:NRQCD}
&d\s_{A+B\rightarrow \jpsi +
X}=\sum_{i,j,n} \int dx_1 dx_2 f_{i/A}(x_1) f_{j/B}(x_2) d{\s}_{i +
j\rightarrow \cc[n] + X_H}\frac{\mopn}{m^{d_n}}\,,
\end{align}
where $m$ is the mass of charm quark, the denominator $m^{d_n}$ has
the same mass dimension as $\mopn$ which insures $d{\s}$ to have the same mass
dimension on both sides, $f$ is parton distribution function and $d{\s}_{i +
	j\rightarrow \cc[n] + X_H}$ are short-distance coefficients to produce a $\cc$ pair with quantum number $n$. LDMEs
$\mopn$ are defined as expectation values of four fermions operators in
vacuum, which can be linear combination of forms as
\begin{align} \label{eq:opdef}
\mo_n = \chi^\dagger \mathcal{K}^{'\dagger}_n \psi
\mathcal{P}_{\jpsi} \psi^\dagger \mathcal{K}_n \chi\,,
\end{align}
where $\psi$ ($\chi$) is field of heavy (anti-)quark in NRQCD
effective field theory~\cite{Lepage:1987gg}, $\mathcal{K}_n$
and $\mathcal{K}^{'\dagger}_n$ contain a color matrix, a spin matrix
and a polynomial of covariant differential operator \textbf{D},
respectively and $\mathcal{P}_{\jpsi}$ is the projection operator of
$\jpsi$ with the form
\begin{align} \mathcal{P}_\jpsi = a^\dagger_{\jpsi}a_{\jpsi} = \sum \limits_{X_{\rm S}} |
\jpsi+X_{\rm S} \rangle \langle \jpsi+X_{\rm S}|\,,
\end{align}
where  $X_{\rm S}$ includes all soft hadrons. 
In principle, one needs infinite number of LDMEs to reproduce QCD
result. However, as each LDME has a definite power counting in $v$,
not all of them are relevant to finite
accuracy~\cite{Bodwin:1994jh}. 
The most relevant LDMEs for $\jpsi$ production are $\mops$,
$\mopa$, $\mopb$ and $\mopj$ ($J=0,1,2$), with
\begin{align}\label{eq:mop}
\begin{split}
\mops=&\langle 0| \chi^{\dagger}\sigma^i\psi
\mathcal{P}_\jpsi \psi^{\dagger}\sigma^i \chi |0\rangle\,,\\
\mopa=&\langle 0| \chi^{\dagger}T^{a}\psi
\mathcal{P}_\jpsi \psi^{\dagger} T^{a}\chi |0\rangle\,,\\
\mopb=&\langle 0| \chi^{\dagger}\sigma^i T^{a}\psi
\mathcal{P}_\jpsi \psi^{\dagger}\sigma^i T^{a}\chi |0\rangle\,,\\
\mopc=&\frac{1}{3}\langle 0| \chi^{\dagger} (-\frac{i}{2}\lrD\cdot
\boldsymbol{\sigma}) T^{a}\psi \mathcal{P}_\jpsi \psi^{\dagger}
(-\frac{i}{2}\lrD\cdot \boldsymbol{\sigma}) T^{a}\chi |0\rangle\,,
\end{split}
\end{align}
where the operator $\lrD$ is defined as $\chi^\dagger \lrD \psi =
\chi^\dagger (\boldsymbol{D} \psi) -
(\boldsymbol{D}\chi^\dagger)\psi$ and the approximate heavy quark spin symmetry ensures $\mopj =
(2J+1) \mopc \left(1 + O(v^2)\right)$. We will also use the
following definition in this paper
\begin{align}\label{eq:mopP}
\begin{split}
\mopP=\sum_{J=0,1,2}\mopj=&\langle 0| \chi^{\dagger}
(-\frac{i}{2}\overleftrightarrow{D}^j \sigma^i) T^{a}\psi
\mathcal{P}_\jpsi \psi^{\dagger}
(-\frac{i}{2}\overleftrightarrow{D}^j\sigma^i) T^{a}\chi |0\rangle.
\end{split}
\end{align}

To consider the first order relativistic corrections, one needs the following relatively $v^2$ suppressed LDMEs~\cite{Fan:2009zq, Xu:2012am},
\begin{align}\label{eq:pop}
\begin{split}
\pops=&\frac{1}{2}\left[ \langle 0| \chi^{\dagger}\sigma^i
(-\frac{i}{2}\lrD)^2\psi
\mathcal{P}_\jpsi \psi^{\dagger}\sigma^i \chi |0\rangle + h.c.\right]\,,\\
\popa=&\frac{1}{2}\left[ \langle 0| \chi^{\dagger}
(-\frac{i}{2}\lrD)^2T^{a}\psi
\mathcal{P}_\jpsi \psi^{\dagger} T^{a}\chi |0\rangle + h.c.\right]\,,\\
\popb=&\frac{1}{2}\left[ \langle 0| \chi^{\dagger}\sigma^i
(-\frac{i}{2}\lrD)^2T^{a}\psi
\mathcal{P}_\jpsi \psi^{\dagger}\sigma^i T^{a}\chi |0\rangle + h.c.\right]\,,\\
\popP=&\frac{1}{2}\left[ \langle 0| \chi^{\dagger}
(-\frac{i}{2}\lrD)^2(-\frac{i}{2}\overleftrightarrow{D}^j \sigma^i)
T^{a}\psi \mathcal{P}_\jpsi \psi^{\dagger}
(-\frac{i}{2}\overleftrightarrow{D}^j \sigma^i) T^{a}\chi |0\rangle
+ h.c. \right]\,,
\end{split}
\end{align}
which have two more \textbf{D}'s comparing with the corresponding LDMEs in
Eqs.~\eqref{eq:mop} and~\eqref{eq:mopP}. At even higher order in $v$, there can be a lot of LDMEs.

To use NRQCD factorization, one needs to calculate short-distance
coefficients $d{\s}_{i + j\rightarrow \cc[n] + X_H}$ in
Eq.~\eqref{eq:NRQCD}. 
We denote the short-distance coefficient as $F(n)$ and $G(n)$ if
the corresponding LDME is $\mopN$ and $\popN$, respectively.

In the rest frame of the $\cc$ pair~\footnote{We call it rest frame in
the rest of this paper for simplicity.}, we denote their momenta in
the amplitude as
\begin{align}
\begin{split}
p_c=&(E,\vq)\,,\\
p_{\cbar}=&(E,-\vq)\,,
\end{split}
\end{align}
where $\vq$ is half of the relative momentum between $\cc$ pair,
$E=\sqrt{m^2+|\vq|^2}$ is half of the invariant mass of $\cc$ pair.
In the complex conjugated amplitude, the relative momentum $\vq^\prime$
can be in principle different from $\vq$. Momentum conservation
gives $p_c+p_{\cbar} = p_c^\prime+p_{\cbar}^\prime$, which results
in $E^\prime=E$ and $|\vq^\prime|=|\vq|$. Therefore, we do not
distinguish $E^\prime$ and $|\vq^\prime|$ from $E$ and $|\vq|$ in
the following. It is convenient to define a ratio
\begin{align}
\b=\frac{|\vq|}{E}\,,
\end{align}
which is also the same in amplitude and complex conjugated
amplitude. In arbitrary frame,
\begin{align}
\begin{split}
p_c=&\frac{1}{2}P+q\,,\\
p_{\cbar}=&\frac{1}{2}P-q\,,
\end{split}
\end{align}
with $P$ is the total momentum of $\cc$ pair and $q$ is boosted from
$(0,\vq)$ in the rest frame. To produce a $\cc[n]$ state, one should
project both color and spin of $\cc$ pair to this definite state.
There are two kinds of color states with projection operators
\begin{subequations}\label{eq:colorProj}
\begin{align}
\mathcal{C}_{1}=&\frac{\d_{ij}}{\sqrt{N_c}}\,,\\
\mathcal{C}_{8}^a=&{\sqrt{2}T^a_{ij}}\,,
\end{align}
\end{subequations}
for singlet and octet respectively. For spin singlet or triplet, one
needs the following projection operators
\begin{subequations} \label{eq:spinProj}
\begin{align}
\Pi_{1}=&\frac{1}{\sqrt{2E}(E+m)}\left(\slashed{p}_{\bar
c}-m\right) \frac{2E-\slashed{P}}{4E} \g^5
\frac{2E+\slashed{P}}{4E}
\left(\slashed{p}_c+m\right)\,,\\
\Pi_{3}^\a=&\frac{1}{\sqrt{2E}(E+m)}\left(\slashed{p}_{\bar
c}-m\right) \frac{2E-\slashed{P}}{4E} \g^\a
\frac{2E+\slashed{P}}{4E} \left(\slashed{p}_c+m\right)\,.
\end{align}
\end{subequations}

Denoting momenta of light partons in initial state and final state as
$k_1$, $k_2$ and $k_3$ respectively, one has Lorentz invariant
Mandelstam variables
\begin{align}
\begin{split}
s=&(k_1+k_2)^2=(P+k_3)^2\,,\\
t=&(k_2-k_3)^2=(P-k_1)^2\,,\\
u=&(k_1-k_3)^2=(P-k_2)^2\,,
\end{split}
\end{align}
with $s+t+u=P^2=4E^2$. In principle, both $F(n)$ and $G(n)$ are
complicated functions of $s,~t,~u$, and $m$. However, in
Refs.~\cite{Fan:2009zq, Xu:2012am}, it was found that there are very
simple relations in large $p_T$ limit (i.e. in the limit that
$p_T^2~\sim~s,~t,~u\gg~m^2$),
\begin{subequations}
\begin{align}
R^{(1)}(\CScSa)=&\left.\frac{G(\CScSa)}{F(\CScSa)}\right|_{p_T\gg
m}=\frac{1}{6}\,,\label{eq:ratio3s11}\\
R^{(1)}(\COaSz)=&\left.\frac{G(\COaSz)}{F(\COaSz)}\right|_{p_T\gg
m}=-\frac{5}{6}\,,\label{eq:ratio1s08}\\
R^{(1)}(\COcSa)=&\left.\frac{G(\COcSa)}{F(\COcSa)}\right|_{p_T\gg
m}=-\frac{11}{6}\,,\label{eq:ratio3s18}\\
R^{(1)}(\COcP)=&\left.\frac{G(\COcP)}{F(\COcP)}\right|_{p_T\gg
m}=-\frac{31}{30}\,,\label{eq:ratio3p8}
\end{align}
\end{subequations}
where the superscript ``$(1)$'' means the ratio of the first order
of relativistic corrections to the lowest order results. Among them,
relation for $\CScSa$ channel is understood in
Ref.~\cite{Ma:2012ex}. In the rest of this paper, we will be
devoted to understand the proportional relations for CO channels and generalize them
to higher orders.

\section{Collinear factorization for $\jpsi$ production}\label{sec:coll}

Production cross section of the $\jpsi$ in the collinear factorization
is given by~\cite{Kang:2014tta,Kang:2014pya}
\begin{align}\label{eq:pqcdfac}
\begin{split}
d\sigma_{A+B\to \jpsi+X}(p) \approx &
\sum\limits_{i,j}f_{i/A}(x_1)f_{j/B}(x_2) \left\{ \sum_{f}
d{\sigma}_{i+j\to f+X}(p_f={\hat P}/z)
\otimes D_{\psi/f}(z,m)\right.\\
& \left. \hspace{-2.5cm}  + \sum_{[\cc(\kappa)]}
d{\sigma}_{i+j\to [\cc(\kappa)]+X}({\hat P}(1\pm\zeta)/2z,{\hat
P}(1\pm\zeta')/2z) \otimes {\cal
D}_{\psi/[\cc(\kappa)]}(z,\zeta,\zeta',m) \right\}\,,
\end{split}
\end{align}
where $p$ is the momentum of the observed $J/\psi$ in the final states. The first (second) term on the right-hand side gives the
contribution of LP (NLP) in $m^2/p_T^2$ expansion. $D_{\psi/f}(z,m)$ is the fragmentation function (FF) for $\jpsi$ from a single parton $f$ with
momentum fraction $z$ and  ${\cal D}_{\psi/[\cc(\kappa)]}(z,\zeta,\zeta',m)$ is the
charm quark-pair FF. Operator
definition of single parton FF can be found in
Refs.~\cite{Braaten:1993rw, Nayak:2005rw, Nayak:2005rt}, and that of
double parton FF can be found in Refs.~\cite{Kang:2014tta,Kang:2014pya}.
The hard-scattering function $d{\sigma}_{i+j\to f+X}(p_f={\hat
P}/z)$ ($d{\sigma}_{i+j\to [\cc(\kappa)]+X}({\hat
P}(1\pm\zeta)/2z,{\hat P}(1\pm\zeta')/2z)$ ) describes the production of
an on-shell light parton (collinear $\cc$ pair).

The pair $[\cc(\kappa)]$ in fragmentation function is massive and,
in general, is off shell, thus gauge link is needed to ensure gauge
invariance. The charm quark pair $[\cc(\kappa)]$ in hard part are massless and moving in the ``$+z$'' direction with
light-cone momentum components ${\hat P}^\mu/z=({\hat P}^+/z,
0,\mathbf{0}_\perp)$, where ${\hat P}^\mu$ is a light like momentum
whose plus component equals to the $\jpsi$'s plus component ${\hat P}^+
= P^+_{\jpsi} = n \cdot {P_{\jpsi}}$. $\kappa$ represents the pair's
color and spin, which has two kinds of color states with the
projection operators defined in Eq.~\eqref{eq:colorProj} for both
fragmentation and hard part, and three kinds of spin states, for
effective axial vector ($a$), vector ($v$) and tensor ($t$)
``currents'', described by relativistic Dirac spin projection
operators
~\cite{Kang:2014tta,Kang:2014pya}
\begin{subequations}
\begin{align}
\widetilde{\mathcal{P}}_a=&\frac{1}{4p^+}\g^+\g_5\,,\\
\widetilde{\mathcal{P}}_v=&\frac{1}{4p^+}\g^+\,,\\
\widetilde{\mathcal{P}}_t^\mu=&\frac{1}{4p^+}\g^+\g_\perp^\mu\,,
\end{align}
\end{subequations}
for FFs and
\begin{subequations}
\begin{align}
\mathcal{P}_a=&\slashed{p}\g^5\,,\\
\mathcal{P}_v=&\slashed{p}\,,\\
\mathcal{P}_t^\mu=&\slashed{p}\g_\perp^\mu\,,
\end{align}
\end{subequations}
for hard parts. The momentum fractions $z$, $\zeta$ and $\zeta'$ are
defined as
\begin{align} \label{eq:zzetadef}
\begin{split}
\hat p_c^\mu &= \frac{1+\zeta}{2z}\, {\hat P}^\mu\, , \,\quad\quad
\hat p_{\bar c}^\mu = \frac{1-\zeta}{2z}\, {\hat P}^\mu\, , \\
\hat p_c^{'\mu} &= \frac{1+\zeta'}{2z}\, {\hat P}^\mu\, , \quad\quad
\hat p_{\bar c}^{'\mu} = \frac{1- \zeta'}{2z}\, {\hat P}^\mu\, .
\end{split}
\end{align}
$z$ measures the fractional momentum of the collinear $\cc$ pair
carried by $\jpsi$ in this leading region, which is the same on both
amplitude side and its complex conjugate side.

Both single parton and double parton FFs can be in principle
determined by fitting experimental data. However, it will be much
powerful if we combine this collinear factorization with NRQCD
factorization. Applying NRQCD factorization to $\jpsi$ FFs, all FFs
can be expressed in terms of some unknown LDMEs
\begin{align}
D_{\psi/f}(z,m)=&\sum_n D_{\cc[n]/f}(z,m) \frac{\mopn}{m^{d_n}}\, ,\\
{\cal D}_{\psi/[\cc(\kappa)]}(z,\zeta,\zeta',m)=&\sum_n {\cal
D}_{\cc[n]/[\cc(\kappa)]}(z,\zeta,\zeta',m) \frac{\mopn}{m^{d_n}}\,
,
\end{align}
Note the difference that $[\cc(\kappa)]$ is a perturbative QCD
state, while $\cc[n]$ is a NRQCD state.

What was shown above is a two steps factorization: first, using
collinear factorization to express the cross section in terms of
hard parts and FFs of a heavy quarkonium; then using NRQCD
factorization to express these FFs in terms of LDMEs. However, it is
more convenient for us to reorganize the factorization as: first,
using NRQCD factorization to express the cross section in terms of
short-distance coefficients and LDMEs; then using collinear
factorization to express the short-distance coefficients in terms of
multiplication of hard parts with FFs of $\cc[n]$. Based on the later
factorization steps, we can study relations between short-distance
coefficients of $\jpsi$ production at large $p_T$ limit for each channel directly. 
Single parton fragmentation functions have been studied extensively~\cite{Braaten:1993mp, Braaten:1994kd, Braaten:1995cj, Ma:1995vi, Qi:2007sf, Hao:2009fa, Bodwin:2014bia, Nejad:2014iba, Nejad:2015oca, Nejad:2015far, Zhang:2017xoj, Feng:2017cjk, MoosaviNejad:2018ukp,Braaten:2000pc, Lee:2005jw, Bodwin:2003wh, Sang:2009zz, Bodwin:2012xc, Gao:2016ihc, MoosaviNejad:2016qdx, Sepahvand:2017gup, Artoisenet:2014lpa, Artoisenet:2018dbs, Feng:2018ulg, Zhang:2018mlo}.
Double parton fragmentation functions have also been calculated to order $\alpha_sv^0$~\cite{Ma:2013yla, Ma:2014eja, Ma:2015yka}.

Recall that we are interested in $\jpsi$ production in $\CScSa$,
$\COcSa$, $\COaSz$ and $\COcP$ channels at
LO in $\a_s$. $\COcSa$ channel is dominated by gluon fragmentation
and $d\s/dp_T^2$ behaves as $p_T^{-4}$ in large $p_T$ limit.
$\COaSz$ and $\COcP$ channels do not have single particle
fragmentation contributions at this order, and their leading
contributions come from $\cc$ pair fragmentation, which behaves as
$p_T^{-6}$. $\CScSa$ channel behaves as $p_T^{-8}$~\cite{Ma:2012ex},
thus it cannot be interpreted in terms of the above formula. In the
following sections, we will discuss the three CO channels one by
one.

\section{Relativistic effect of color-octet $\jpsi$ production}\label{sec:CORC}

\subsection{Relativistic effect in $\COcSa$ channel}\label{sec:3s18}

The short-distance coefficient for producing a $\cc[\COcSa]$ in large
$p_T$ limit can be factorized as
\begin{align}\label{eq:3s1fac}
d\s_{i+j\to \cc[\COcSa]+k}=\int_0^1 dz ~d\s_{i+j\to g+k}(z)
D_{\cc[\COcSa]/g}(z,m)\,,
\end{align}
where $i$, $j$ and $k$ denote various light partons. At LO in
$\a_s$, we can denote $D_{\cc[\COcSa]/g}(z,m) = \overline{D}_{\cc[\COcSa]/g}(m) \d(1-z)$, which results in
\begin{align}
d\s_{i+j\to \cc[\COcSa]+k}=d\s_{i+j\to g+k}(1) \overline
D_{\cc[\COcSa]/g}(m)\,.
\end{align}
Then
we get the ratio
\begin{align}
R(\COcSa)=\frac{d\s_{i+j\to \cc[\COcSa]+k}}{d\s^{(0)}_{i+j\to
\cc[\COcSa]+k}}=\frac{\overline D_{\cc[\COcSa]/g}(m)}{\overline
D^{(0)}_{\cc[\COcSa]/g}(m)}\,,
\end{align}
where the superscript ``$(0)$'' denotes keeping the relative
momentum to the lowest order. Therefore, to calculate $R(\COcSa)$,
one needs only calculate $D_{\cc[\COcSa]/g}(z,m)$. Expanding the
above equation to NLO of relative momentum, we get the expression of
$R^{(1)}(\COcSa)$ in Eq.~\eqref{eq:ratio3s18}
\begin{align}
R^{(1)}(\COcSa)=\frac{\overline D^{(1)}_{\cc[\COcSa]/g}(m)}{\overline
D^{(0)}_{\cc[\COcSa]/g}(m)}\,.
\end{align}

\begin{figure}[htb!]
 \begin{center}
 \includegraphics[width=0.30\textwidth]{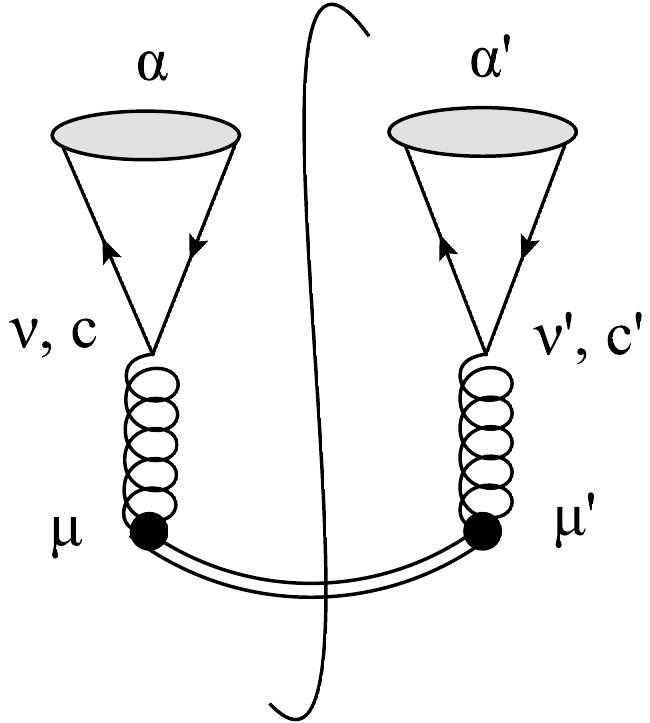}
  \caption[$\jpsi$]{The cut diagram of gluon fragmentation to $\cc[\COcSa]$.
   \label{fig:gFrag}}
 \end{center}
\end{figure}
The cut diagram of gluon fragmentation to $\cc[\COcSa]$ is shown in
Fig.~\ref{fig:gFrag}. The amplitude is
\begin{align}
A^{\a\mu}=\text{Tr}\left[\mathcal{C}_{8}^h\Pi_3^\a(-i g_s T^c
\g^\nu) \right] \frac{-ig^{\mu\nu}}{P^2}\,,
\end{align}
where $\a$ ($h$) is the polarization (color) index of $\cc[\COcSa]$.
Considering that $P^\mu$ and $P^\a$ can be set to zero here, we
have
\begin{align}
A^{\a\mu}=\frac{g_s\d^{hc}}{2\sqrt{E}E^2(E+m)}\left[E(E+m)g^{\a\mu}+q^\a
q^\mu\right]\,.
\end{align}
Then using Eq.~\eqref{eq:S24} to project the above amplitude to
S-wave by the replacement $q^\a q^\mu \to \Pi^{\a\mu}$, we get
\begin{align}
A^{\a\mu}=\frac{g_s\d^{hc}(2E+m)}{6\sqrt{E}E^2}g^{\a\mu}\,.
\end{align}
FF is achieved by squaring the amplitude and summing/averaging over
quantum numbers
\begin{align}
D_{\cc[\COcSa]/g}(z,m)=\frac{1}{2}\frac{1}{3}\frac{1}{(N_c^2-1)^2} P^{\a\a^\prime}
d^{\mu\mu^\prime} A^{\a\mu} A^{\dagger \a^\prime \mu^\prime}
\d(1-z)=\frac{g_s^2}{96}\frac{\left(\frac{1+2\Delta}{3}\right)^2}{\Delta^5m^3}\d(1-z)\,,
\end{align}
with $\Delta=E/m$, $P^{\a\b}=-g^{\a\b}+\frac{P^\a P^{\b}}{P^2}$ and
$d^{\a\b}=-g^{\a\b}+\frac{P^\a n^\b+n^\a P^\b}{P\cdot n}-\frac{P^2
n^\a n^\b}{(P\cdot n)^2}$. This result agrees with the calculation in Ref.~\cite{Ma:2017xno}. Therefore, we get
\begin{align}
R(\COcSa)=\frac{\left(\frac{1+2\Delta}{3}\right)^2}{\Delta^5}\,.
\end{align}
The expansion of $R(\COcSa)$ as powers of $\d=\abq^2/m^2$ is
\begin{align}\label{eq:3s18}
R(\COcSa)=1 - \frac{11}{6} \d + \frac{191}{72} \d^2 - \frac{167}{48}
\d^3 + \cdots \,,
\end{align}
where the second term indeed reproduces the result in
Eq.~\eqref{eq:ratio3s18} of NLO relativistic correction in NRQCD
framework. This result tells us that the proportional relation for $\COcSa$ channel holds to all orders in velocity expansion. We will see that this property holds also for the other two channels.

To give an
estimation of the convergence of velocity expansion in this part, we choose an average value for $\d$ in Eq.\eqref{eq:3s18}. The expansions of $R(\COcSa)$ are
listed in Tab.~\ref{table:3s18}, where $\d$ is
chosen as $0.3$ for $\jpsi$ and $0.1$ for $\Upsilon$. From
Tab.~\ref{table:3s18}, we find that after a few corrections, the
expansion can converge to its exact result soon. Obviously, the
convergence of $\Upsilon$ system is much faster than that of the $\jpsi$
system.
\begin{table}
\begin{center}
\begin{tabular}{|c|c|c|c|c|c|c|c|}
  \hline
  $n$&~~~0~~~ & ~~~1~~~ & ~~~2~~~ &~~~3~~~&~~~4~~~& ~~$\cdots$~~ & ~~$\infty$~~\\
  \hline
  $\left.\sum_{i=0}^n \d^i R^{(i)}(\COcSa)\right|_{\d=0.3}$&1 & 0.450 & 0.689 & 0.595 & 0.630 & $\cdots$ & 0.620\\
  \hline
  $\left.\sum_{i=0}^n \d^i R^{(i)}(\COcSa)\right|_{\d=0.1}$&1 & 0.817 & 0.843 & 0.840 & 0.840 & $\cdots$ & 0.840\\
  \hline
\end{tabular}
\end{center}
\caption[]{Perturbative expansion of $R(\COcSa)$. $\d=\abq^2/m^2$ is
chosen to be $0.3$ and $0.1$ in the second and third row,
respectively.
   \label{table:3s18}}
\end{table}

\subsection{Relativistic effect in $\COaSz$ channel}\label{sec:1s08}

A $\cc$ pair which fragments to $\cc[\COaSz]$ at LO in $\a_s$ must
be in CO. Then, there are three possible channels, $a^{[8]}$,
$v^{[8]}$ and $t^{[8]}$. The hard part $i+j\to [\cc(t^{[8]})] + k$
at LO in $\a_s$ are helicity suppressed, whose contributions are
at higher power in $m^2/p_T^2$, thus $t^{[8]}$ channel can be
neglected. ${\cal D}_{\cc[\COaSz]/[\cc(v^{[8]})]}$ vanishes at LO in
$\a_s$ because there are only three independent vectors but a $\g^5$ exists in the trace of gamma matrices. All in all, the
short-distance coefficient for producing a $\cc[\COaSz]$ in large
$p_T$ limit can be factorized as
\begin{align}\label{eq:1s0fac}
d\s_{i+j\to \cc[\COaSz]+k}=\int dz d\zeta d\zeta' ~d\s_{i+j\to
[\cc(a^{[8]})]+k}(z,\zeta,\zeta') {\cal
D}_{\cc[\COaSz]/[\cc(a^{[8]})]}(z,\zeta,\zeta',m)\,.
\end{align}

\begin{figure}[htb!]
 \begin{center}
 \includegraphics[width=0.30\textwidth]{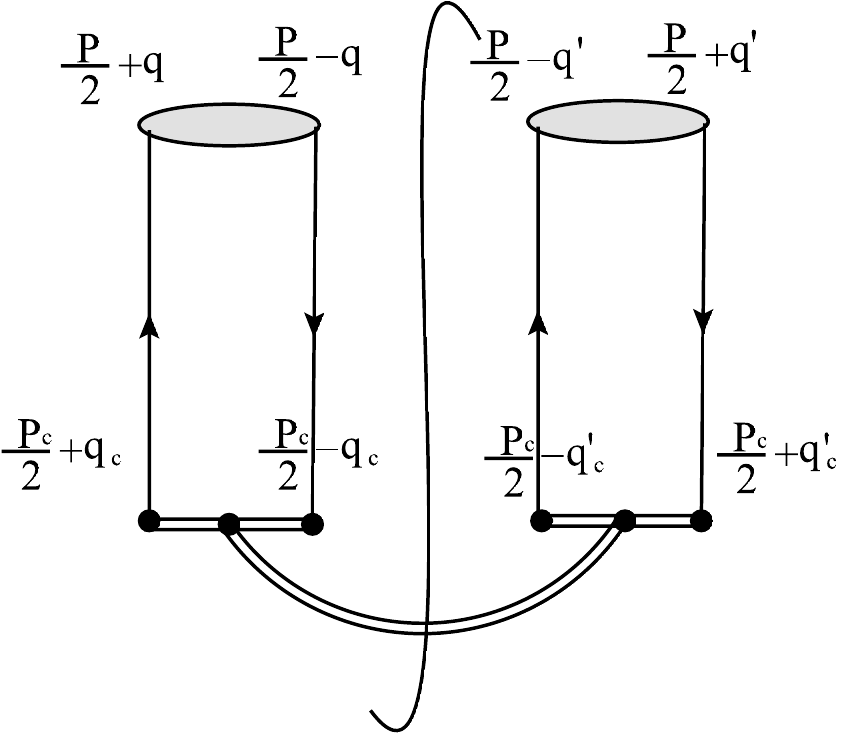}
  \caption[$\jpsi$]{The cut diagram of a $\cc$ pair in
perturbative QCD fragmentate to a $\cc$ pair in NRQCD.
   \label{fig:ccFrag}}
 \end{center}
\end{figure}

Let's first calculate the fragmentation function ${\cal
D}_{\cc[\COaSz]/[\cc(a^{[8]})]}$. The cut diagram for a perturbative QCD $\cc$ pair 
fragments to a NRQCD $\cc$ pair is shown in
Fig.~\ref{fig:ccFrag}. The amplitude for ${\cal
D}_{\cc[\COaSz]/[\cc(a^{[8]})]}$ is
\begin{align}
\begin{split}
A&=\int
\text{Tr}\left[\mathcal{C}_8^h\Pi_1\mathcal{C}_8^{h_c}\tilde{\mathcal{P}}_a
\right]
(2\pi)^4\d^4(\frac{P}{2}+q-\frac{P_c}{2}-q_c)2\d(\zeta-\frac{2q_c^+}{P_c^+})\frac{d^4q_c}{(2\pi)^4}\\
&=\text{Tr}\left[\mathcal{C}_8^h\Pi_1\mathcal{C}_8^{h_c}\tilde{\mathcal{P}}_a
\right] 2\d(\zeta-\frac{2q^+}{P^+})\,,
\end{split}
\end{align}
where the relation $P_c=P$ guaranteed by momentum conservation has
been used. It is straightforward to get
\begin{align}
A=\frac{-2m}{(2E)^{3/2}}\d^{hh_c}\d(\zeta-\frac{2q^+}{P^+})\,.
\end{align}
Squaring the amplitude and summing/averaging over quantum numbers,
we get
\begin{align}
{\cal
D}_{\cc[\COaSz]/[\cc(a^{[8]})]}(z,\zeta,\zeta',m)=\frac{4}{(N_c^2-1)^2}
 \frac{8m^2}{(2E)^{3}}\d(\zeta-\frac{2q^+}{P^+})
\d(\zeta'-\frac{2q'^+}{P^+})\d(1-z)\,.
\end{align}

It was found in Ref.~\cite{Kang:2014pya} that hard parts have the
behavior $d\s_{i+j\to [\cc(a^{[8]})]+k}(z,\zeta,\zeta')=C_a^{ijk} +
(\text{terms odd in $\zeta$ or $\zeta'$})$, where $C_a^{ijk}$ is independent of $\zeta$ and $\zeta'$. Thus the
Eq.~\eqref{eq:1s0fac} becomes
\begin{align}
\begin{split}
&d\s_{i+j\to \cc[\COaSz]+k}=\int dz d\zeta d\zeta' ~(C_a^{ijk} +
\text{terms odd in $\zeta$
or $\zeta'$})\\
&\qquad\qquad\times
\frac{m^2}{2 (1-\zeta^2)(1-\zeta'^2)(2E)^{3}}\d(\zeta-\frac{2q^+}{P^+})
\d(\zeta'-\frac{2q'^+}{P^+})\d(1-z)\\
=&\left. (C_a^{ijk} + \text{terms odd in $\zeta$ or
$\zeta'$})
\frac{m^2}{2 (1-\zeta^2)(1-\zeta'^2)(2E)^{3}}
\right|_{\zeta=\frac{2q^+}{P^+},
\zeta'=\frac{2q'^+}{P^+}}\\
=&\left. C_a^{ijk}
\frac{m^2}{2 (1-\zeta^2)(1-\zeta'^2)(2E)^{3}}\right|_{\zeta=\frac{2q^+}{P^+},
\zeta'=\frac{2q'^+}{P^+}}\,,
\end{split}
\end{align}
where in the last equation  we suppress terms odd in $\zeta$
(or $\zeta'$) which do not contribute to S-wave. Keep to lowest
order, one gets
\begin{align}
\begin{split}
&d\s_{i+j\to \cc[\COaSz]+k}^{(0)}=C_a^{ijk}
\frac{m^2}{2 (2m)^{3}}\,,
\end{split}
\end{align}
and
\begin{align}
R(\COaSz)=\frac{m^3}{(1-\zeta^2)(1-\zeta'^2)E^{3}}\,.
\end{align}
Note that we have still not projected the result to S-wave in the
above expression. This projection can be done using
Eq.~\eqref{eq:Szeta3} by replacing $(1-\zeta'^2)^{-1}$ and $(1-\zeta^2)^{-1}\to \b^{-1}
\text{arctanh}(\b)$. The final result is
\begin{align}
\begin{split}
R(\COaSz)=&\left(\frac{m}{E}\right)^3
\left(\frac{\text{arctanh}(\b)}{\b}\right)^2\\
=&\frac{\left(\text{arctanh}(\sqrt{\frac{\d}{1+\d}})\right)^2}{\d\sqrt{1+\d}}\\
=&1 - \frac{5}{6} \d + \frac{259}{360} \d^2 - \frac{3229}{5040} \d^3
+ \cdots \,.
\end{split}
\end{align}
where the second term in the last line indeed reproduces the result
in Eq.~\eqref{eq:ratio1s08} of NLO relativistic correction for
$\COaSz$ channel in NRQCD framework, while higher order terms here
are new. As an estimate of convergence of perturbative expansion,
choosing $\d=0.3$ and $\d=0.1$, the expansions of $R(\COaSz)$ are
listed in Tab.~\ref{table:1s08}, where a good convergence similar as
$R(\COcSa)$ is found.
\begin{table}
\begin{center}
\begin{tabular}{|c|c|c|c|c|c|c|}
  \hline
  $n$&~~~0~~~ & ~~~1~~~ & ~~~2~~~ &~~~3~~~& ~~$\cdots$~~ & ~~$\infty$~~\\
  \hline
  $\left.\sum_{i=0}^n \d^i R^{(i)}(\COaSz)\right|_{\d=0.3}$&1 & 0.750 & 0.815 & 0.797 & $\cdots$ & 0.801\\
  \hline
  $\left.\sum_{i=0}^n \d^i R^{(i)}(\COaSz)\right|_{\d=0.1}$&1 & 0.917 & 0.924 & 0.923 & $\cdots$ & 0.923\\
  \hline
\end{tabular}
\end{center}
\caption[]{Perturbative expansion of $R(\COaSz)$. $\d=\abq^2/m^2$ is
chosen to be $0.3$ and $0.1$ in second and third row, respectively.
   \label{table:1s08}}
\end{table}

\subsection{Relativistic effect in $\COcP$ channel}\label{sec:3pj8}

Similar analysis as that for $\COaSz$ channel gives that the
short-distance coefficient for producing a $\cc[\COcP]$ in large
$p_T$ limit can be factorized as
\begin{align}\label{eq:3pfac}
\begin{split}
d\s_{i+j\to \cc[\COcP]+k}=&\int dz d\zeta d\zeta' ~d\s_{i+j\to
[\cc(a^{[8]})]+k}(z,\zeta,\zeta') {\cal
D}_{\cc[\COcP]/[\cc(a^{[8]})]}(z,\zeta,\zeta',m)\\
&+d\s_{i+j\to [\cc(v^{[8]})]+k}(z,\zeta,\zeta') {\cal
D}_{\cc[\COcP]/[\cc(v^{[8]})]}(z,\zeta,\zeta',m)\\
=&d\s_a+ d\s_v\,.
\end{split}
\end{align}

\subsubsection{Axial-vector  channel}
The amplitude for ${\cal D}_{\cc[\COcP]/[\cc(a^{[8]})]}$ is
\begin{align}
\begin{split}
A_a^\a&=\int
\text{Tr}\left[\mathcal{C}_8^h\Pi_3^\a\mathcal{C}_8^{h_c}\tilde{\mathcal{P}}_a
\right]
(2\pi)^4\d^4(\frac{P}{2}+q-\frac{P_c}{2}-q_c)2\d(\zeta-\frac{2q_c^+}{P_c^+})\frac{d^4q_c}{(2\pi)^4}\\
&=\text{Tr}\left[\mathcal{C}_8^h\Pi_3^\a\mathcal{C}_8^{h_c}\tilde{\mathcal{P}}_a
\right]
2\d(\zeta-\frac{2q^+}{P^+})\\
&=\frac{2i\e^{Pnq\a}}{(2E)^{3/2}P^+}\d^{hh_c}\d(\zeta-\frac{2q^+}{P^+})\,.
\end{split}
\end{align}
Squaring the amplitude and summing/averaging over quantum numbers,
we get
\begin{align}
\begin{split}
{\cal
D}_{\cc[\COcP]/[\cc(a^{[8]})]}(z,\zeta,\zeta',m)=&\frac{1}{9}\frac{1}{(N_c^2-1)^2}P^{\a\a'}A_a^\a
A_a^{\dag\a'}\d(1-z)\\
& \hspace{-1cm} = \frac{-q\cdot q' - E^2 \zeta \zeta'
}{18 (2E)^{3}}\d(\zeta-\frac{2q^+}{P^+})
\d(\zeta'-\frac{2q'^+}{P^+})\d(1-z)\,,
\end{split}
\end{align}
which is odd in $q$ and $q'$ if we do not consider delta functions.
As only terms with odd number of $q$ and $q'$ contribute to P-wave
and hard parts have the behavior  $d\s_{i+j\to [\cc(a^{[8]})]+k}(z,\zeta,\zeta')=C_a^{ijk}
+ (\text{terms odd in $\zeta$ or $\zeta'$})$, we have
\begin{align}
\begin{split}
d\s_a=&\int dz d\zeta d\zeta'~C_a^{ijk}\frac{-q\cdot q' -  E^2 \zeta
\zeta' }{18 (1-\zeta^2)(1-\zeta'^2)(2E)^{3}}\d(\zeta-\frac{2q^+}{P^+})
\d(\zeta'-\frac{2q'^+}{P^+})\d(1-z)\\
=&\left. C_a^{ijk}\frac{-q\cdot q' -  E^2 \zeta \zeta'
}{18 (1-\zeta^2)(1-\zeta'^2)(2E)^{3}}\right|_{\zeta=\frac{2q^+}{P^+},
\zeta'=\frac{2q'^+}{P^+}}\,.
\end{split}
\end{align}
Using Eq.~\eqref{eq:zetaP} to project the above expression to
P-wave, we get
\begin{align}
\begin{split}
d\s_a=&C_a^{ijk}\frac{\abq^2}{18 (2E)^{3}}\left\{ (3\Delta_1^2+\Delta_2
\Delta_3) - (\Delta_1^2+\Delta_2 \Delta_3) \right\} =
C_a^{ijk}\frac{\abq^2}{9 (2E)^{3}} \Delta_1^2 \,,
\end{split}
\end{align}
and
\begin{align}
\begin{split}
d\s_a^{(0)}= C_a^{ijk}\frac{\abq^2}{9 (2m)^{3}} \,.
\end{split}
\end{align}
Therefore,
\begin{align}
\begin{split}
R_a(\COcP)=&\left(\frac{m}{E}\right)^3\Delta_1^2\\
=&\frac{9}{4} \frac{\sqrt{1+\d}}{\d^2} \left[-\frac{1}{\sqrt{\d(1+\d)}}\text{arctanh}(\sqrt{\frac{\d}{1+\d}}) +1\right]^2\\
=&1 - \frac{11}{10} \d + \frac{1521}{1400} \d^2 - \frac{8803}{8400}
\d^3 + \cdots \,.
\end{split}
\end{align}

\subsubsection{Vector  channel}
The amplitude for ${\cal D}_{\cc[\COcP]/[\cc(v^{[8]})]}$ is
\begin{align}
\begin{split}
A_v^\a&=\int
\text{Tr}\left[\mathcal{C}_8^h\Pi_3^\a\mathcal{C}_8^{h_c}\tilde{\mathcal{P}}_v
\right]
(2\pi)^4\d^4(\frac{P}{2}+q-\frac{P_c}{2}-q_c)2\d(\zeta-\frac{2q_c^+}{P_c^+})\frac{d^4q_c}{(2\pi)^4}\\
&=\text{Tr}\left[\mathcal{C}_8^h\Pi_3^\a\mathcal{C}_8^{h_c}\tilde{\mathcal{P}}_v
\right]
2\d(\zeta-\frac{2q^+}{P^+})\\
&=\frac{-2}{P^+\sqrt{2E}}\left( E\, n^\a + \frac{q^+q^\a}{E+m}
\right)\d^{hh_c}\d(\zeta-\frac{2q^+}{P^+})\,.
\end{split}
\end{align}
Squaring the amplitude and summing/averaging over quantum numbers,
we get
\begin{align}
\begin{split}
{\cal
D}_{\cc[\COcP]/[\cc(v^{[8]})]}(z,\zeta,\zeta',m)=&\frac{1}{9}\frac{1}{(N_c^2-1)^2}P^{\a\a'}A_v^\a
A_v^{\dag\a'}\d(1-z)\\
& \hspace{-3cm} = \frac{ 1 - \frac{E}{E+m} (\zeta^2+\zeta'^2) -
\frac{q\cdot q'}{(E+m)^2} \zeta\zeta'
}{18 (2E)^{3}}E^2\d(\zeta-\frac{2q^+}{P^+})
\d(\zeta'-\frac{2q'^+}{P^+})\d(1-z)\,,
\end{split}
\end{align}
which is even in $q$ and $q'$ if we do not consider delta functions.
As only terms with odd number of $q$ and $q'$ contribute to P-wave
and hard parts have behaviors $d\s_{i+j\to
[\cc(v^{[8]})]+k}(z,\zeta,\zeta')=C_v^{ijk} \zeta\zeta'+ (\text{terms even
in $\zeta$ or $\zeta'$})$~\cite{Kang:2014pya}, where $C_v^{ijk}$ is independent of $\zeta$ and $\zeta'$, we have
\begin{align}
\begin{split}
d\s_v=&\int dz d\zeta d\zeta'~C_v^{ijk} \zeta\zeta'\frac{ 1 -
\frac{E}{E+m} (\zeta^2+\zeta'^2) - \frac{q\cdot q'}{(E+m)^2}
\zeta\zeta'
}{18 (1-\zeta^2)(1-\zeta'^2)(2E)^{3}}E^2\d(\zeta-\frac{2q^+}{P^+})
\d(\zeta'-\frac{2q'^+}{P^+})\d(1-z)\\
=& \left. C_v^{ijk} \zeta\zeta'\frac{ 1 - \frac{E}{E+m} (\zeta^2+\zeta'^2)
- \frac{q\cdot q'}{(E+m)^2} \zeta\zeta'
}{18 (1-\zeta^2)(1-\zeta'^2)(2E)^{3}}E^2\right|_{\zeta=\frac{2q^+}{P^+},
\zeta'=\frac{2q'^+}{P^+}}\\
=& \frac{C_v^{ijk}}{144 E} \left\{ \frac{m-E}{m+E} \frac{\zeta
\zeta'}{(1-\zeta^2)(1-\zeta'^2)} + \frac{E}{m+E} \zeta \zeta'
\left( \frac{1}{1-\zeta^2} + \frac{1}{1-\zeta'^2}\right) \right.\\
& \left. \left. -\frac{q\cdot q'}{(m+E)^2} \left[
\frac{1}{(1-\zeta^2)(1-\zeta'^2)} - \frac{1}{1-\zeta^2}
-\frac{1}{1-\zeta'^2} +1 \right] \right\}
\right|_{\zeta=\frac{2q^+}{P^+}, \zeta'=\frac{2q'^+}{P^+}} \,.
\end{split}
\end{align}
Using Eq.~\eqref{eq:zetaP} to project the above expression to
P-wave, we get
\begin{align}
\begin{split}
d\s_v=&\frac{\abq^2}{18}\frac{C_v^{ijk}}{8E} \left\{ \frac{m-E}{E^2(m+E)}
(\Delta_1^2 + \Delta_2 \Delta_3) +
\frac{2}{E(m+E)} (\Delta_1+\Delta_2) \right.\\
& \left. +\frac{1}{(m+E)^2} \left[ (3\Delta_1^2 + \Delta_2 \Delta_3)
- 2 (3\Delta_1 + \Delta_2) +3 \right] \right\} \,.
\end{split}
\end{align}
Considering that
\begin{align}
\begin{cases}
\Delta_1=1+O(\d)\,,\\
\Delta_2=\frac{2\d}{5}+O(\d^2)\,,\\
\Delta_3=2+O(\d)\,,
\end{cases}
\end{align}
we get
\begin{align}
\begin{split}
d\s_v^{(0)}= C_v^{ijk}\frac{\abq^2}{18 (2m)^{3}} \,.
\end{split}
\end{align}
Therefore,
\begin{align}
\begin{split}
R_v(\COcP)=&\frac{m^3}{E} \left\{ \frac{m-E}{E^2(m+E)} (\Delta_1^2 +
\Delta_2 \Delta_3) +
\frac{2}{E(m+E)} (\Delta_1+\Delta_2) \right.\\
& \left. +\frac{1}{(m+E)^2} \left[ (3\Delta_1^2 + \Delta_2
\Delta_3) - 2 (3\Delta_1 + \Delta_2) +3 \right] \right\}\\
=&1 - \frac{9}{10} \d + \frac{1069}{1400} \d^2 - \frac{5549}{8400}
\d^3 + \cdots \,.
\end{split}
\end{align}

\subsubsection{Summation of the two channels}

Finally, we get
\begin{align}
\begin{split}
R(\COcP)=&\frac{d\s_a+d\s_v}{d\s_a^{(0)}+d\s_v^{(0)}}
=\frac{R_a(\COcP)+ \frac{C_v^{ijk}}{2C_a^{ijk}} R_v(\COcP)}{1+\frac{C_v^{ijk}}{2C_a^{ijk}}}.
\end{split}
\end{align}
In Ref.~\cite{Kang:2014pya}, it is interesting to find that $C_a^{ijk}=C_v^{ijk}$, which results in 
\begin{align}
\begin{split}
R(\COcP)=&\frac{2R_a(\COcP)+R_v(\COcP)}{3}\\
=&1 - \frac{31}{30} \d + \frac{4111}{4200} \d^2 - \frac{4631}{5040}
\d^3 + \cdots \,,
\end{split}
\end{align}
where the second term in the last line indeed reproduces the result
in Eq.~\eqref{eq:ratio3p8} of NLO relativistic correction for
$\COcP$ channel in NRQCD framework, while higher order terms here
are new. As an estimate of convergence of perturbation expansion,
choosing $\d=0.3$ and $\d=0.1$, the expansions of $R(\COcP)$ are
listed in Tab.~\ref{table:3p8}, where the convergence is good.
\begin{table}
\begin{center}
\begin{tabular}{|c|c|c|c|c|c|c|}
  \hline
  $n$&~~~0~~~ & ~~~1~~~ & ~~~2~~~ &~~~3~~~& ~~$\cdots$~~ & ~~$\infty$~~\\
  \hline
  $\left.\sum_{i=0}^n \d^i R^{(i)}(\COcP)\right|_{\d=0.3}$&1 & 0.690 & 0.778 & 0.753 & $\cdots$ & 0.759\\
  \hline
  $\left.\sum_{i=0}^n \d^i R^{(i)}(\COcP)\right|_{\d=0.1}$&1 & 0.897 & 0.906 & 0.906 & $\cdots$ & 0.906\\
  \hline
\end{tabular}
\end{center}
\caption[]{Perturbative expansion of $R(\COcP)$. $\d=\abq^2/m^2$ is
chosen to be $0.3$ and $0.1$ in second and third row, respectively.
   \label{table:3p8}}
\end{table}

\section{Summary and outlook}\label{sec:summary}

By combining NRQCD factorization and collinear factorization for
heavy quarkonium production, we calculate the relativistic
correction for $\jpsi$ hadron production to all orders in $v^2$ at
large $p_T$ limit. As large $p_T$ data are very important to determine CO LDMEs in NRQCD, our calculation should be useful for this purpose. It is interesting to find that the ratio
of relativistic correction contribution to the leading contribution  $R(n)$ is independent
of kinematic variables for all production channels, which generalizes the finding in Ref.~\cite{Xu:2012am} to all orders in $v^2$ expansion. Specifically,
for $\COcSa$ channel, relative momentum does not flow into hard part
after the factorization in Eq.~\eqref{eq:3s1fac}, thus kinematics
dependence in $R(\COcSa)$ cancels between denominator and numerator.
For $\COaSz$ channel, relative momentum still flows into hard part
in terms of $\zeta$ and $\zeta'$ even after the factorization in
Eq.~\eqref{eq:1s0fac}. However, to produce a S-wave $\cc$ pair, only
a relative momentum independent kinematic configuration $C_a^{ijk}$ in
hard part contributes, which cancels between denominator and
numerator of $R(\COaSz)$. The $\COcP$ channel is even complicated
because two terms contribute in the factorization formula in
Eq.~\eqref{eq:3pfac}. Fortunately, kinematic configurations for the
two terms ($C_a^{ijk}$ and $C_v^{ijk}$) are identical, thus they can be canceled
in $R(\COcP)$.

By expanding $R(n)$ according to powers of $\d=\abq^2/m^2$,
we find $O(\d)$ terms reproduce large $p_T$ results in
Ref.~\cite{Xu:2012am}. By setting $\d=0.3$ or $\d=0.1$, we find a good convergence for $R(n)$ expansion. 

Note that, the calculation in this paper only covers one series of relativistic correction terms. Another series of relativistic correction terms caused by soft gluons emission are also important for $\jpsi$ hadroproduction, which can be taken into account by the soft gluon factorization~\cite{Ma:2017xno}. Our calculation is also a start point for the use of soft gluon factorization framework.

\begin{acknowledgments}
We thank J.W. Qiu and H. Zhang for useful discussions. This work is supported in part by the National Natural Science Foundation
of China (Grant No. 11875071) and the China Postdoctoral Science Foundation under Grant No.2018M631234.
\end{acknowledgments}

\appendix
\renewcommand{\theequation}{\thesection\arabic{equation}}

\section{Average over direction of relative momentum}\label{sec:app}

In this appendix, we give explicit expressions that project short
distance coefficients to definite orbital angular momentum state.
Define projection operator $F_{lm}$, which projects amplitude to
state with orbital angular momentum $l$ and polarization $m$. In the
rest frame of the intermediate $\cc$ pair and parameterizing the
relative momentum as $q^\mu=(0,\vq)=\abq(0,\sin \th \cos \phi, \sin
\th \sin \phi,\cos \th)$, $F_{lm}$ acts on amplitude $A(\vq)$ gives
\begin{align}
F_{lm}(A(\vq))=\sqrt{\frac{(2l+1)!!}{4\pi (l!)}} \int{ d\Omega
Y^m_l(\th,\phi) A(\vq) },
\end{align}
where $Y^m_l(\th,\phi)$ are spherical harmonics functions. If there is
no L-S coupling, we can use the following projection operator to get
unpolarized differential cross section
\begin{align}
F_{l}(M(\vq,\vq^\prime))= \sum_{m=\pm l, \pm (l-1), \cdots, 0}
\frac{(2l+1)!!}{4\pi (l!)} \int d\Omega d\Omega^\prime
Y^m_l(\th,\phi) \left(Y^m_l(\th^\prime,\phi^\prime)\right)^*
M(\vq,\vq^\prime),
\end{align}
which acts on the squared amplitudes. In this work we consider only
S-wave ($l=0$) and P-wave ($l$=1), explicit spherical harmonics
functions of which are
\begin{align}
Y^0_0(\th,\phi)=\frac{1}{\sqrt{4\pi}}, \qquad
Y^0_1(\th,\phi)=\sqrt{\frac{3}{4\pi}}\cos \th, \qquad
Y^{\pm1}_1(\th,\phi)=\mp\sqrt{\frac{3}{8\pi}}\sin{\th} e^{\pm i
\phi}. \nonumber
\end{align}
Once the projected results in the rest frame are calculated, results
in arbitrary frame can be easily obtained using Lorentz covariance.
Specifically, one needs to do the following replacement
\begin{align}
\abq^2 &\to -q^2,\\
k^0 &\to \frac{P\cdot k}{P^2}P^\mu,\\
\d^{ij}&\to \Pi^{\mu\nu}=-g^{\mu\nu}+\frac{P^\mu P^\nu}{P^2},
\end{align}
where $k^0$ is the ``0"-component of a momentum $k$.

\subsection{S-wave}\label{sec:appS}

For S-wave, projection operator gives
\begin{align}\label{eq:S}
F_{0}(A(\vq))=\frac{1}{\sqrt{4\pi}} \int{ d\Omega
Y^0_0(\th,\phi)A(\vq) }=\frac{1}{4\pi}\int_{-1}^{1} d \cos\th
\int_0^{2\pi}d\phi A(\vq),
\end{align}
which satisfies
\begin{align}
F_{0}(1)=1.
\end{align}
Because of parity symmetry, $F_{0}$ acts on product of odd number of
$\vq$ vanishes
\begin{align}
F_{0}(\prod_{j=1}^{2N+1}q^{i_j})=0.
\end{align}
$F_{0}$ acts on two $\vq$ gives
\begin{align}
\begin{split}
&F_{0}(\vq\cdot\vec{\mathbf{k}}_1 \vq\cdot\vec{\mathbf{k}}_2)\\
=&\frac{\abq^2}{4\pi}\int_{-1}^{1} d \cos\th
\int_0^{2\pi}d\phi \left(k_{1x}\sin\th\cos\phi+k_{1y}\sin\th\sin\phi+k_{1z}\cos\th\right)\\
& \qquad\qquad\qquad\qquad\qquad\times\left(k_{2x}\sin\th\cos\phi+k_{2y}\sin\th\sin\phi+k_{2z}\cos\th\right)\\
=&\frac{\abq^2}{4\pi}\int_{-1}^{1} d \cos\th
\int_0^{2\pi}d\phi \left(k_{1x}k_{2x}\sin\th^2\cos\phi^2+k_{1y}k_{2y}\sin\th^2\sin\phi^2+k_{1z}k_{2z}\cos\th^2\right)\\
=&\frac{\abq^2}{3} \left(k_{1x}k_{2x}+k_{1y}k_{2y}+k_{1z}k_{2z}\right)\\
=&\frac{\abq^2}{3}\vec{\mathbf{k}}_1\cdot\vec{\mathbf{k}}_2,
\end{split}
\end{align}
which means
\begin{align}\label{eq:S23}
F_{0}(q^{i}q^j)=\frac{\abq^2}{3}\delta^{ij}.
\end{align}
In arbitrary frame, it gives
\begin{align}\label{eq:S24}
F_{0}(q^{\mu}q^\nu)=\frac{-q^2}{3}\Pi^{\mu\nu}.
\end{align}
Expression in Eqs.~\eqref{eq:S23} and \eqref{eq:S24} can be
generalized to any even number of $\vq$. Notice that there is no
vector in $F_{0}(\prod_{j=1}^{2N}q^{i_j})$ after integration, thus
its result must be a symmetric tensor containing only terms like
$\delta^{ij}$. Therefore we have
\begin{align}\label{eq:S2Npre}
F_{0}(\prod_{j=1}^{2N}q^{i_j})\propto \{[\d]^N\}^{i_1\cdots i_{2N}},
\end{align}
where $\{[\d]^N\}^{i_1\cdots i_{2N}}$ means a tensor combination
symmetric under $i_1,~\cdots,~i_{2N}$, and each term in the
combination has $N$ delta functions. For example, $\{[\d]^2\}^{i_1
i_2 i_3
i_{4}}=\delta^{i_1i_2}\delta^{i_3i_4}+\delta^{i_1i_3}\delta^{i_2i_4}+\delta^{i_1i_4}\delta^{i_2i_3}$.
It is easy to find that $\{[\d]^N\}^{i_1\cdots i_{2N}}$ contains
$(2N-1)!!$ terms. Contract both sides of Eq.~\eqref{eq:S2Npre} with
$\d^{i_1i_2}\d^{i_3i_4}\cdots \d^{i_{2N-1}i_{2N}}$, one can
determine its proportionality factor to be
$\frac{\abq^{2N}}{(2N+1)!!}$. Thus we get
\begin{align}\label{eq:S2N3}
F_{0}(\prod_{j=1}^{2N}q^{i_j})= \frac{\abq^{2N}}{(2N+1)!!}
\{[\d]^N\}^{i_1\cdots i_{2N}},
\end{align}
and
\begin{align}\label{eq:S2N4}
F_{0}(\prod_{j=1}^{2N}q^{\mu_j})= \frac{(-q^2)^{N}}{(2N+1)!!}
\{[\Pi]^N\}^{\mu_1\cdots \mu_{2N}},
\end{align}

In this work, we also need to calculate $F_{0}(\frac{1}{1-\zeta^2})$
with $\zeta=\frac{2q\cdot n}{P\cdot n}$. It is convenient to
calculate it in the rest frame and choose $n$ to only have $t$ and $z$ directions.
We get
\begin{align}\label{eq:Szeta3}
F_{0}(\frac{1}{1-\zeta^2})= \frac{1}{4\pi}\int_{-1}^{1} d \cos\th
\int_0^{2\pi}d\phi\frac{1}{1-\left(\frac{2\abq \cos
\th}{2E}\right)^2}=\b^{-1} \text{arctanh}(\b),
\end{align}
where
\begin{align}
\b=\frac{\abq}{E}.
\end{align}

\subsection{P-wave}\label{sec:appP}

For P-wave, projection operator gives
\begin{align}\label{eq:P}
F_{1m}(A(\vq))=\sqrt{\frac{3}{4\pi}} \int{ d\Omega
Y^m_1(\th,\phi)A(\vq) }=\sqrt{\frac{3}{4\pi}}\int_{-1}^{1} d \cos\th
\int_0^{2\pi}d\phi Y^m_1(\th,\phi) A(\vq).
\end{align}
Parity symmetry in this case results in that $F_{1m}$ acts on
product of even number of $\vq$ vanishes
\begin{align}
F_{1m}(\prod_{j=1}^{2N}q^{i_j})=0,
\end{align}
which can be checked explicitly in Eq.~\eqref{eq:P}. $F_{1m}$ acts
on one $\vq$ gives
\begin{align}\label{eq:P13}
\begin{split}
F_{1m}(\vq\cdot\vec{\mathbf{k}})=&\sqrt{\frac{3}{4\pi}}\int_{-1}^{1}
d \cos\th
\int_0^{2\pi}d\phi Y^m_1(\th,\phi) \abq \left(k_{x}\sin\th\cos\phi+k_{y}\sin\th\sin\phi+k_{z}\cos\th\right)\\
=&\abq\left(\d_{m,0}k_z + \d_{m,1}\frac{-k_x-ik_y}{\sqrt{2}}+
\d_{m,-1}\frac{k_x-ik_y}{\sqrt{2}}\right)\\
\equiv&\abq \vec{\mathbf{\epsilon}}_m\cdot\vec{\mathbf{k}},
\end{split}
\end{align}
where $\vec{\mathbf{\epsilon}}_m$ are polarization vectors
\begin{align}
\vec{\mathbf{\epsilon}}_0=(0,0,1),~~~\vec{\mathbf{\epsilon}}_{\pm
1}=\frac{1}{\sqrt{2}}(\mp 1,-i,0),
\end{align}
which satisfy the orthonormality
\begin{align}
\sum_{i=1}^3\epsilon_m^i
(\epsilon_{m'}^*)^i=\d_{mm'},~~~\text{and}~~~\sum_{m=0,\pm1}\epsilon_m^i
(\epsilon_{m}^*)^{i'}=\d^{ii'},
\end{align}
Therefore, by summing over $m$ for the squared amplitude, we get
\begin{align}
\begin{split}
&\sum_{m=0,\pm1}F_{1m}(\vq\cdot\vec{\mathbf{k}})\left[F_{1m}(\vq^\prime\cdot\vec{\mathbf{k}}^\prime)\right]^*
=\abq^2\vec{\mathbf{k}}\cdot\vec{\mathbf{k}}^\prime,
\end{split}
\end{align}
where the fact that $|\vq^\prime|=\abq$ is used. Then we have
\begin{align}\label{eq:P113}
\begin{split}
F_1(q^iq^{\prime
j})=&\sum_{m=0,\pm1}F_{1m}(q^i)\left[F_{1m}(q^{\prime j})\right]^*
=\abq^2\d^{ij}.
\end{split}
\end{align}
This method also applies when there are more $\vq$ or $\vq^\prime$.
For example, one can easily get
\begin{align}\label{eq:P313}
\begin{split}
F_{1}(q^{i_1}q^{i_2}q^{i_3}q^{\prime i_4}) =\frac{\abq^4}{5}\left(
\delta^{i_1i_2}\delta^{i_3i_4}+\delta^{i_1i_3}\delta^{i_2i_4}+\delta^{i_1i_4}\delta^{i_2i_3}
\right).
\end{split}
\end{align}

We then calculate $F_{1m}(\frac{q^i}{1-\zeta^2})$. Choosing the same
convention as for the S-wave case, we have
\begin{align}\label{eq:Pzeta13}
\begin{split}
&F_{1m}(\frac{\vq\cdot\vec{\mathbf{k}}}{1-\zeta^2})\\
=&\sqrt{\frac{3}{4\pi}}\int_{-1}^{1}
d \cos\th \int_0^{2\pi}d\phi Y^m_1(\th,\phi)
\frac{\abq}{1-\left(\frac{2\abq \cos
\th}{2E}\right)^2} \left(k_{x}\sin\th\cos\phi+k_{y}\sin\th\sin\phi+k_{z}\cos\th\right)\\
=&\abq\left\{ \frac{3}{\b^2}\left[\b^{-1}\text{arctanh}(\b)-1\right]
\d_{m,0}k_z + \frac{3}{2\b^2}\left[(\b-\b^{-1})\text{arctanh}(\b)+1
\right]  \right.\\
&\qquad \qquad\qquad\qquad \qquad\qquad \left.
\times\left(\d_{m,1}\frac{-k_x-ik_y}{\sqrt{2}}+
\d_{m,-1}\frac{k_x-ik_y}{\sqrt{2}}\right)\right\}\\
=&\abq\left\{\Delta_1 \vec{\mathbf{\epsilon}}_m\cdot\vec{\mathbf{k}}
+ \Delta_2 \d_{m,0}k_z \right\},
\end{split}
\end{align}
where $\Delta_1$ and $\Delta_2$ are defined as
\begin{align}
\Delta_1&=\frac{3}{2\b^2}\left[(\b-\b^{-1})\text{arctanh}(\b)+1
\right],\\
\Delta_2&=\frac{3}{2\b^2}\left[(3\b^{-1}-\b)\text{arctanh}(\b)-3
\right].
\end{align}
Summing over $m$, we get
\begin{align}
\begin{split}
&\sum_{m=0,\pm1}F_{1m}(\frac{\vq\cdot\vec{\mathbf{k}}}{1-\zeta^2})
\left[F_{1m}(\frac{\vq^\prime\cdot\vec{\mathbf{k}}^\prime}{1-\zeta^{\prime2}})\right]^*\\
=&\abq^2\left\{ \Delta_1^2 \vec{\mathbf{k}} \cdot
\vec{\mathbf{k}}^\prime + \Delta_2(2\Delta_1+\Delta_2) k_z
k_z^\prime\right\}\\
=&\abq^2\left\{ \Delta_1^2 \vec{\mathbf{k}} \cdot
\vec{\mathbf{k}}^\prime + \Delta_2\Delta_3 \frac{P^2
\vec{\mathbf{k}} \cdot \vec{\mathbf{n}} \vec{\mathbf{k}}^\prime
\cdot \vec{\mathbf{n}} }{(P\cdot n)^2}\right\},
\end{split}
\end{align}
where
\begin{align}
\Delta_3&=2\Delta_1+\Delta_2=\frac{3}{2\b^2}\left[(\b+\b^{-1})\text{arctanh}(\b)-1
\right].
\end{align}
Therefore
\begin{align}\label{eq:Pzz113}
\begin{split}
F_1(\frac{q^iq^{\prime
j}}{(1-\zeta^2)(1-\zeta^{\prime2})})=\abq^2\left\{ \Delta_1^2
\d^{ij} + \Delta_2\Delta_3 \frac{P^2 n^i n^j}{(P\cdot n)^2}\right\}.
\end{split}
\end{align}
Similarly, we have
\begin{align}\label{eq:Pz113}
\begin{split}
F_1(\frac{q^iq^{\prime j}}{1-\zeta^2})=\abq^2\left\{ \Delta_1
\d^{ij} + \Delta_2 \frac{P^2 n^i n^j}{(P\cdot n)^2}\right\}.
\end{split}
\end{align}
Generalizing Eqs.~\eqref{eq:P113}, \eqref{eq:Pzz113} and
\eqref{eq:Pz113} to arbitrary frame and contracting indexes, we get
the following explicit expressions that are used in this work,
\begin{subequations}\label{eq:zetaP}
\begin{align}
F_1(-q\cdot q^\prime)&=\abq^23,\\
F_1(\frac{-q\cdot q^\prime}{1-\zeta^2})&=\abq^2(3\Delta_1+\Delta_2),\\
F_1(\frac{-q\cdot q^\prime}{(1-\zeta^2)(1-\zeta^{\prime2})})&=\abq^2(3\Delta_1^2+\Delta_2 \Delta_3),\\
F_1(\zeta\zeta^\prime)&=\abq^2\frac{1}{E^2},\\
F_1(\frac{\zeta\zeta^\prime}{1-\zeta^2})&=\abq^2\frac{1}{E^2}(\Delta_1+\Delta_2),\\
F_1(\frac{\zeta\zeta^\prime}{(1-\zeta^2)(1-\zeta^{\prime2})})&=\abq^2\frac{1}{E^2}(\Delta_1^2+\Delta_2
\Delta_3)\,.
\end{align}
\end{subequations}


\providecommand{\href}[2]{#2}\begingroup\raggedright\endgroup

\end{document}